\documentclass[12pt]{article} %Set document type and font size.

\usepackage{helvet}

\newcommand*{\myEXPfont}{\fontfamily{cmr}\selectfont}

\usepackage{graphicx} %Allows you to insert pdf and jpg images into your files.
\usepackage{mathtools}
\usepackage{amsmath}
\usepackage{multirow}
\usepackage[export]{adjustbox}
\usepackage[left=1in, right=1in, top=1in, bottom=1in]{geometry} %Sets the page margins as required.
\usepackage{booktabs}
\usepackage{rotating}
\usepackage{rotfloat}
\usepackage{float}
\usepackage{subfig}
\usepackage{url}
\usepackage{pdfpages}
\usepackage{rotating}
\usepackage[doublespacing]{setspace} %Double-space your text as required.
\usepackage{csvsimple}
\usepackage{array}
\usepackage{wrapfig}
\usepackage[utf8]{inputenc}

\usepackage{cite}
\usepackage{tabularx}
\usepackage{float}
\usepackage{verbatim}

\usepackage{amsthm}                % better theorem environments
\usepackage{amssymb}

\usepackage{multirow}

\usepackage{caption}
\usepackage{amsfonts}
\usepackage{amsmath} 
\usepackage{tikz,float}
\usetikzlibrary{positioning,shapes.geometric}

\usepackage{xcolor}

\usepackage{setspace}

\DeclareMathOperator{\E}{\mbox{{\myEXPfont E}}}

\usepackage{mathrsfs}

\makeatletter
\newcommand*{\indep}{%
  \mathbin{%
    \mathpalette{\@indep}{}%
  }%
}
\newcommand*{\nindep}{%
  \mathbin{%                   % The final symbol is a binary math operator
    \mathpalette{\@indep}{\not}% \mathpalette helps for the adaptation
  }%
}
\newcommand*{\@indep}[2]{%
  \sbox0{$#1\perp\m@th$}%        box 0 contains \perp symbol
  \sbox2{$#1=$}%                 box 2 for the height of =
  \sbox4{$#1\vcenter{}$}%        box 4 for the height of the math axis
  \rlap{\copy0}%                 first \perp
  \dimen@=\dimexpr\ht2-\ht4-.2pt\relax
  \kern\dimen@
  {#2}%
  \kern\dimen@
  \copy0 %                       second \perp
} 
\makeatother

\usepackage{xr-hyper}

\definecolor{forestgreen}{RGB}{34,139,34}
\usepackage{hyperref}
\hypersetup{
    colorlinks=true,
    linkcolor=magenta,
    filecolor=magenta, 
    citecolor=forestgreen,      
    urlcolor=blue
}

\usepackage{sectsty}
\sectionfont{\large}

\usepackage{layout}

\usepackage{geometry}

\newcolumntype{C}[1]{>{\centering\arraybackslash}p{#1}}

\usepackage{authblk}

\usepackage{afterpage}
\usepackage{lscape}
\usepackage{graphicx}

\usepackage{datetime}
\def\paperversionmajor{11}
\def\paperversionminor{0}

\makeatletter
\newcommand*{\addFileDependency}[1]{% argument=file name and extension
  \typeout{(#1)}
  \@addtofilelist{#1}
  \IfFileExists{#1}{}{\typeout{No file #1.}}
}
\makeatother

\newcommand*{\myexternaldocument}[1]{%
    \externaldocument{#1}%
    \addFileDependency{#1.tex}%
    \addFileDependency{#1.aux}%
}

\myexternaldocument{appendix}

\usepackage{amsthm}

\raggedright
\begin{document}

\title{The role of assignment in defining and identifying causal effects in randomized trials}

\author[1-3]{Issa J. Dahabreh}
\author[1-2]{Lawson Ung}
\author[1-3]{Miguel A. Hern\'an}
\author[1,2,4]{Yu-Han Chiu}

\affil[1]{CAUSALab, Harvard T.H. Chan School of Public Health, Boston, MA}
\affil[2]{Department of Epidemiology, Harvard T.H. Chan School of Public Health, Boston, MA}
\affil[3]{Department of Biostatistics, Harvard T.H. Chan School of Public Health, Boston, MA}
\affil[4]{Department of Epidemiology, Brown University School of Public Health, Providence, RI, 02903}

\maketitle{}

 \thispagestyle{empty}

\noindent
\textbf{Type of manuscript:} Original Research Article

\noindent
\textbf{Address for correspondence:} Dr. Issa J. Dahabreh, Department of Epidemiology, Harvard T.H. Chan School of Public Health, 677 Huntington Ave., room 816c, Boston, MA 02115; email: \href{mailto:idahabreh@hsph.harvard.edu}{idahabreh@hsph.harvard.edu}; phone: +1 (617) 495‑1000.

\noindent
\textbf{Running head:} Role of assignment in defining treatment effects

\noindent
\textbf{Sources of financial support:} 
This work was supported by Patient-Centered Outcomes Research Institute (PCORI) award ME-2021C2-22365, National Library of Medicine (NLM) award R01LM013616, and National Heart, Lung, and Blood Institute (NHLBI) award R01HL136708.

\noindent
\textbf{Data access:} No datasets are used in this paper.

\clearpage
\vspace*{0.6in}

\setcounter{page}{1}

\begin{abstract}
\linespread{1.5}\selectfont
In randomized trials, the per-protocol effect, that is, the effect of being assigned a treatment strategy and receiving treatment according to the assigned strategy, is sometimes thought to reflect the effect of the treatment strategy itself, without intervention on assignment. Here, we argue by example that this is not necessarily the case. We examine a causal structure for a randomized trial where these two causal estimands -- the per-protocol effect and the effect of the treatment strategy -- are not equal, and where their corresponding identifying observed data functionals are not the same, but both require information on assignment for identification. Our example highlights the conceptual difference between the per-protocol effect and the effect of the treatment strategy, the conditions under which these causal estimands are equal, and suggests that in some cases their identification requires information on assignment, even when assignment is randomized. Furthermore, both per-protocol effects and effects of treatment may be unidentifiable without information on treatment assignment, unless one makes additional assumptions -- informally, that assignment does not affect the outcome except through treatment (i.e., an exclusion-restriction assumption), and that assignment is not a confounder of the treatment-outcome association conditional on other variables in the analysis. Our analyses suggest a need to more clearly define the role of assignment when specifying causal effects of interest in randomized trials, which has implications for identification, analysis methods, and the interpretation of trial results.
\end{abstract}

\clearpage

%%%%%%%%%%%%%%%%%%%%%%%%%%%%%%%%%%%%%%%%%%%%%%%%%%%%%%%%%%%%%%%%%%%%%%%%%%%%%%
\section*{INTRODUCTION}
%%%%%%%%%%%%%%%%%%%%%%%%%%%%%%%%%%%%%%%%%%%%%%%%%%%%%%%%%%%%%%%%%%%%%%%%%%%%%%

In randomized trials, the per-protocol effect is often defined as the effect of receiving the assigned treatment strategies as specified in the trial protocol \cite{hernan2012beyond, hernan2017per}. Therefore, the per-protocol effect encompasses both the effect of following a treatment strategy and any potential effect of the process by which the strategy is assigned. For example, some randomized trials use ``blinded'' assignment so that trial participants remain unaware of the treatment strategy they were assigned to. Furthermore, in most but not all randomized trials, participants are aware that assignment is random; that knowledge may influence adherence to the assigned strategy and outcomes, in ways that are separate from the treatment strategy itself.  It is possible, then, that for the same treatment strategies, the per-protocol effect could vary between trials with different assignment processes (e.g., blinded vs. non-blinded, or with knowledge vs. not of the randomness in the assignment process). This has practical implications for the interpretation of trial results when the process of recommending treatments in practice is materially different from the process of assigning treatments in experimental settings (e.g., blinding and randomization are not commonly used when making treatment recommendations in clinical practice \cite{dahabrehadherence}).

Here, we examine the definition of the per-protocol effect and of the effect of the treatment strategy in a setting of point (time-fixed) treatments and an outcome measured at the end of follow-up. A sharper distinction between these causal estimands, both of which may depend on the assignment process, may be useful for transparently communicating research goals and methods in applied work. Furthermore, we show that identification of per-protocol effects may require information on assignment, even when assignment is randomized. Our findings suggest that per-protocol effects and the effect of a treatment strategy may vary between randomized trials that use different approaches to assign participants to different treatment strategies.

%%%%%%%%%%%%%%%%%%%%%%%%%%%%%%%%%%%%%%%%%%%%%%%%%%%%%%%%%%%%%%%%%%%%%%%%%%%%%%
\section*{STUDY DESIGN \& DATA} \label{section_design_data}
%%%%%%%%%%%%%%%%%%%%%%%%%%%%%%%%%%%%%%%%%%%%%%%%%%%%%%%%%%%%%%%%%%%%%%%%%%%%%%

\paragraph{Study design:} We consider a marginally randomized trial that compares two treatment strategies, each involving a point (non-time-varying) treatment, on an outcome measured at the end of follow-up.  We focus on point treatments and assume that assignment and treatment are binary. Our arguments, however, readily generalize to strategies that require treatment to be sustained over time, and cases where treatment assignment and the received treatment strategy have more than two levels. 

It is well-appreciated that aspects of trial design -- the use of well-defined, protocol-specified interventions; randomized treatment assignment; and non-zero assignment probability to each treatment of interest -- support the identification of causal effects in the trial \cite{robins1986, hernan2006estimating, rubin2007design}. We will formalize these notions after describing the data collected in the trial and introducing our causal model.

\paragraph{Data:} As commonly done in trial analyses, we model the trial data as a simple random sample from an (infinite) population underlying the trial \cite{robins1988confidence, robins2002covariance}. For trial participants, we collect information on assignment to the treatment strategy $Z$; post-assignment covariates $X$, measured after treatment assignment but before receiving treatment; the received treatment $A$; unmeasured covariates $U$; and the outcome measured at the end of the study $Y$ (continuous, count, or binary). Here, the post-assignment covariates $X$ reflect factors that may be affected by assignment, may influence adherence to the assigned strategy, and may be common causes of treatment and the outcome. Such variables may be collected in trials with a delay between randomization and the first administration of treatment, or trials where patient status can change over short periods of time. We use them here to make our methodological arguments under a simple causal structure, but these arguments extend naturally to trials of treatment strategies that are sustained over time, where information on post-assignment covariates is more commonly collected. To simplify exposition, we do not consider information on baseline (pre-assignment) covariates. Such data are not necessary to identify the intention-to-treat effect in marginally randomized trials. Furthermore, in our examples, baseline covariate data are also not necessary to identify the per-protocol effect or the effect of treatment. 

\paragraph{Simplifying assumptions:} We make some additional simplifying assumptions to focus on essential points about the identifiability of different causal estimands. First, we assume no losses to follow-up, even though the issues we discuss also apply to studies with incomplete follow-up or other reasons for censoring. Second, we do not consider effects of trial participation because our goal is learning about the population underlying the randomized trial; issues of generalizability and transportability of causal effects arise when using trials to learn about clinically relevant target populations \cite{dahabreh2019extending_eje, dahabreh2019identification}, but we defer these issues to future work. Last, we assume positivity of the joint distribution of $(Z,X,A)$, such that the joint density $f(Z = z, X = x, A = a)>0$ for all $z,x,a$ values. This positivity condition is stronger than necessary for many of the identification results we provide below, but it does considerably simplify the presentation. We discuss identification under weaker positivity conditions in the Appendix (Section 5). For instance, our results also apply when the probability of assignment to a particular treatment in the trial is zero (e.g., when non-adherent trial participants seek treatments that are different from the ones randomly assigned in the trial).

%%%%%%%%%%%%%%%%%%%%%%%%%%%%%%%%%%%%%%%%%%%%%%%%%%%%%%%%%%%%%%%%%%%%%%%%%%%%%%
\section*{CAUSAL MODEL \& ESTIMANDS} \label{section_causal_model_estimands}
%%%%%%%%%%%%%%%%%%%%%%%%%%%%%%%%%%%%%%%%%%%%%%%%%%%%%%%%%%%%%%%%%%%%%%%%%%%%%%

\paragraph{Counterfactual outcomes and causal model:} We use counterfactual (potential) outcomes \cite{splawaneyman1990, rubin1974, robins1986, robins2000d} to define causal estimands and discuss identifiability conditions. We adopt a finest fully randomized causally interpretable structured tree graph (FFRCISTG) causal model \cite{robins1986, richardson2013single}. In this model, counterfactuals induced by intervening on any variable are considered well-defined, and consistency holds for all these interventions. 

Specifically, we consider the following variables well-defined: $X^{z}$, for $z=0,1$, the counterfactual post-assignment covariates under intervention to set assignment $Z$ to $z$; $A^{z}$, for $z=0,1$, the counterfactual treatment under intervention to set assignment $Z$ to $z$; $Y^{z}$, for $z=0,1$, the counterfactual outcome under intervention to set assignment $Z$ to $z$; $Y^{z,a}$, for $z=0,1$ and $a=0,1$, the counterfactual outcomes under joint intervention to set assignment $Z$ to $z$ and treatment $A$ to $a$; and $Y^{a}$, for $a=0,1$, the counterfactual outcomes under intervention to set treatment $A$ to $a$.

Our causal model includes assumptions of consistency for all possible interventions \cite{richardson2013single}. That said, we only invoke a subset of these consistency conditions; specifically, the following: if $Z=z$, then $X^{z} = X$, $A^{z} = A$, and $Y^{z} = Y$; if $Z=z$ and $A=a$, then $X^{z,a} = X$ and $Y^{z,a} = Y$; and if $A=a$, then $Y^{a} = Y$. 

\paragraph{Causal estimands:} We define three types of causal estimands that are commonly of interest in a randomized trial. Often, the main estimand of interest is the intention-to-treat effect, which is defined as the contrast between counterfactual outcomes under different assignments, without separate intervention on treatment, $\E[Y^{z=1} - Y^{z=0}]$. 

In the presence of non-adherence to assignment \cite{sheiner1995intention, sheiner2002intent}, which in our setting means that $Z \neq A$ for at least some individuals, we might be interested in the per-protocol effect, that is, the effect of receiving the assigned treatment strategy. Here, that effect is defined using the contrast between counterfactual outcomes under joint intervention to set assignment $Z$ to $z$ and treatment $A$ to the assigned treatment $a = z$, that is, $\E[Y^{z=1, a=1} - Y^{z=0, a=0}]$. 

We may also be interested in the effect of the treatment strategy without intervention on assignment. In our setting, that effect is defined  using the contrast between counterfactual outcomes under different treatment strategy, $\E[Y^{a =1} - Y^{a = 0}]$. Henceforth, we refer to this causal estimand simply as the effect of treatment, because, in our time-fixed setting, the treatment strategy is fully implemented at the time of its receipt. 

We note that all three estimands are \emph{potentially} trial-specific because each depends on the type of assignment. This dependence is explicit in the definition of the intention-to-treat effect and the per-protocol effect (both of which include the superscript $z$ indicating intervention on assignment), but is also true for the effect of treatment. Re-writing the components of the effect of the treatment strategy as $\E[Y^a] = \E[Y^{Z, a}] = \sum_{z} \E[Y^{Z, a} | Z = z] \Pr[Z = z] = \sum_{z} \E[Y^{z, a} | Z = z] \Pr[Z = z]$ reminds us that the potential outcomes under intervention on treatment can potentially depend on the assignment process and its effect on the outcome, as they occur in the population under study.   

In the next section, we introduce two causal structures that highlight differences between the three effects defined above, and examine issues related to their identifiability. The causal structures were chosen to elucidate the conditions under which, in a randomized trial, the per-protocol effect and the effect of treatment are equal and when they are not. For completeness, we provide a short discussion on additional causal structures -- elaborated further in Appendix (Sections 1, 2, and 7) -- which extend the results reported in the main text. 

%%%%%%%%%%%%%%%%%%%%%%%%%%%%%%%%%%%%%%%%%%%%%%%%%%%%%%%%%%%%%%%%%%%%%%%%%%%%%%
\section*{EXAMPLE CAUSAL STRUCTURE \# 1} \label{section_causal_model_estimands}
%%%%%%%%%%%%%%%%%%%%%%%%%%%%%%%%%%%%%%%%%%%%%%%%%%%%%%%%%%%%%%%%%%%%%%%%%%%%%%

\begin{figure}[H]
    \caption{Causal structure \# 1. (A) Directed acyclic graph; (B) SWIG under intervention to set assignment $Z$ to $z$; (C) SWIG under intervention to set assignment $Z$ to $z$ and treatment $A$ to $a$; and (D) SWIG under intervention to set treatment $A$ to $a$.}
    \centering
    \includegraphics[width=1\linewidth]{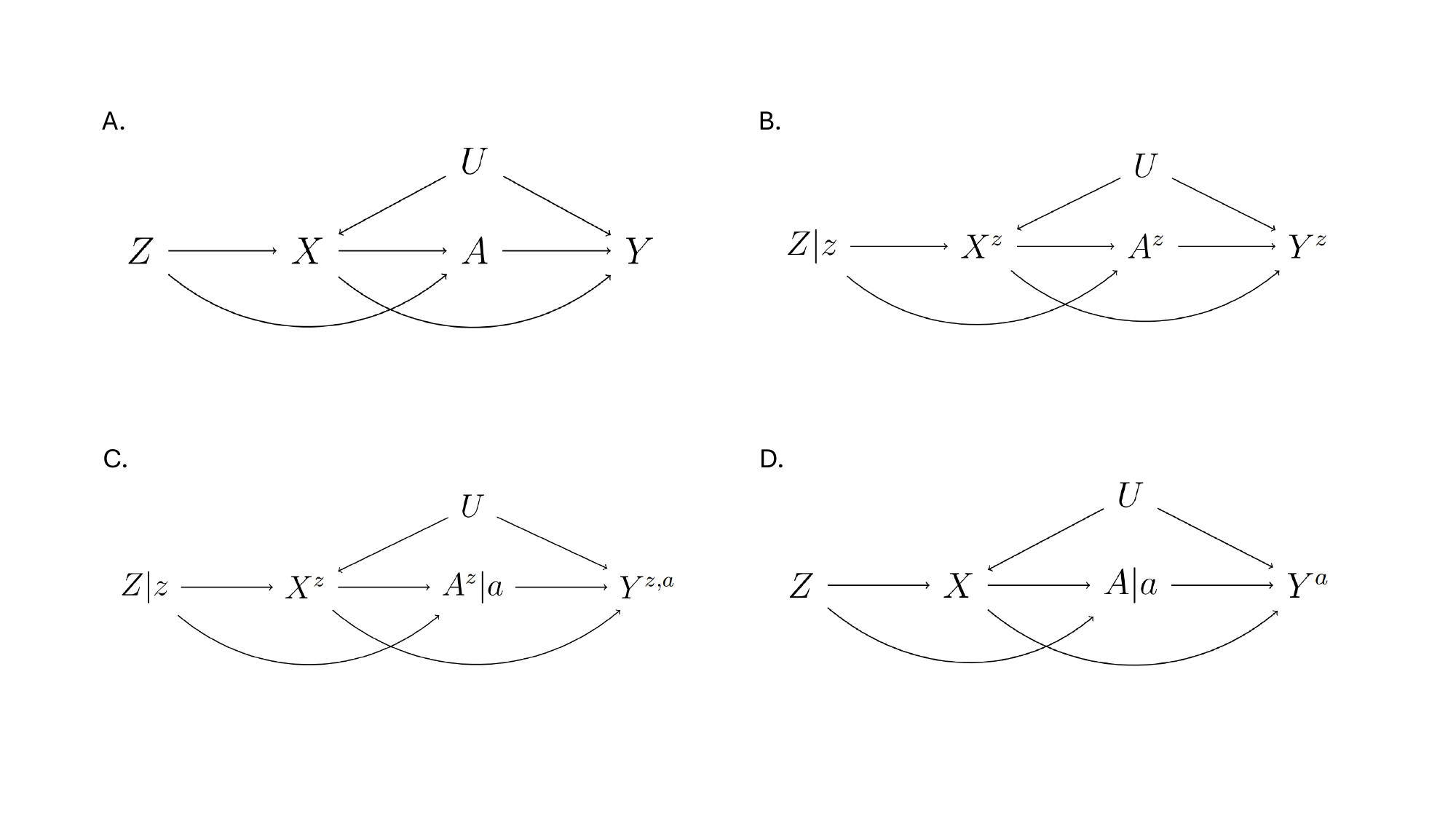}
    \label{fig:DAG1}
\end{figure}

The causal directed acyclic graph (DAG) presented in Figure \ref{fig:DAG1} represents the causal structure for the trial population. Several features of this structure are noteworthy. First, treatment assignment $Z$ and the outcome $Y$ do not share any common causes, measured or unmeasured, reflecting the marginally randomized design of the trial. Second, treatment assignment $Z$ affects the outcome through post-assignment covariates $X$ and through treatment $A$; these effects are represented by the $Z \rightarrow X \rightarrow Y$, $Z \rightarrow A \rightarrow Y$, and $Z \rightarrow X \rightarrow A \rightarrow Y$ directed paths. Third, the unmeasured covariates $U$ operate as common causes of treatment $A$ and the outcome $Y$, but their effect on $A$ is entirely mediated by $X$, that is to say there is an open path (fork) $A \leftarrow X \leftarrow  U \rightarrow Y$, but $U \rightarrow X \rightarrow A$ is the only open directed path from $U$ to $A$ and there is no path from $U$ to $A$ that does not intersect $X$ (an assumption similar to this is often necessary to identify per-protocol effects, e.g., using g-methods \cite{hernan2024causal}). Last, there are no effects of assignment $Z$ on the outcome that are not mediated by the received treatment $A$ or the measured post-assignment covariates $X$; for example there is no direct $Z \rightarrow Y$ path. 

%structure 1, swigs 
\paragraph{Representing interventions on assignment and treatment:} To examine the identifiability of different causal estimands under the DAG of Figure \ref{fig:DAG1}, we use single world intervention graphs (SWIGs) \cite{richardson2013single} to represent intervention on assignment $Z$ (Figure \ref{fig:DAG1}B); joint intervention on assignment $Z$ and treatment $A$ (Figure \ref{fig:DAG1}C); and intervention on treatment $A$ alone (Figure \ref{fig:DAG1}D). In the main text we present identification results using g-formula (nested expectation) expressions \cite{robins1986}; in the Appendix (Section 4) we provide equivalent weighting re-expressions of the identifying functionals. 

% structure 1, ITT
\paragraph{Identification of the intention-to-treat effect:} Under the SWIG in Figure \ref{fig:DAG1}B the independence $Y^{z}\indep Z$ holds. It is well-known \cite{hernan2024causal} that this independence condition, together with the consistency and positivity conditions, suffices to identify the potential outcome mean under intervention to set assignment $Z$ to $z$, $\E[Y^{z}]$, with $$\gamma(z) \equiv \E[Y|Z = z], \mbox{ for } z=0,1,$$ and the intention to treat effect, $\E[Y^{z=1} - Y^{z=0}]$, with
\begin{equation*}
    \delta_{\gamma} \equiv \gamma(z=1) - \gamma(z=0).
\end{equation*}

% structure 1, PP
\paragraph{Identification of the per-protocol effect:} Under the SWIG of Figure \ref{fig:DAG1}C, the independence conditions $Y^{z,a}\indep Z$ and $Y^{z,a} \indep A^z | (X^z, Z)$ hold. As we show in the Appendix (Sections 1-3), these independence conditions, together with the consistency and positivity conditions, suffice to identify the potential outcome mean under intervention to set assignment $Z$ to $z$ and treatment $A$ to $a$, $\E[Y^{z,a}]$, with $$\phi(z,a) \equiv \E \big[  \E[Y|Z, X, A = a ] \big | Z = z \big], \mbox{ for } z=0,1 \mbox{ and } a=0,1,$$ and the per-protocol effect, $\E[Y^{z=1,a=1} - Y^{z=0,a=0}]$ with 
\begin{equation*}
    \delta_{\phi} \equiv \phi(z=1,a=1) - \phi(z=0,a=0).
\end{equation*}           

% structure 1, EOT
\paragraph{Identification of the effect of treatment:} Under the SWIG of Figure \ref{fig:DAG1}D, the independence condition $Y^{a}\indep A | (Z, X)$ holds. As we show in the Appendix (Sections 1-3), this independence condition, together with the consistency and positivity conditions, suffice to identify the potential outcome mean under intervention to set treatment $A$ to $a$, $\E[Y^{a}]$, with $$\chi(a) \equiv \E \big[  \E[Y| Z, X, A = a ] \big], \mbox{ for } a=0,1,$$ and the effect of treatment, $\E[Y^{a=1} - Y^{a=0}]$, with 
\begin{equation*}
    \delta_{\chi} \equiv \chi(a=1) - \chi(a=0).
\end{equation*}

% structure 1, relationship between stat estimands
\paragraph*{Is the per-protocol effect the same as the effect of treatment?} The DAG of Figure \ref{fig:DAG1}A has a directed path from assignment $Z$ to the outcome $Y$ that does not intersect with treatment $A$, that is, the $Z \rightarrow X \rightarrow Y$ path. Because of this path, we have no reason to believe that $Y^{z,a}$ is equal to $Y^a$, reflecting the effect of assignment on the outcome not mediated by treatment. Combining this observation with the identification results above, we conclude that we do not in general expect the per-protocol effect $\E[Y^{z=1,a=1} - Y^{z=0,a=0}]$ and the effect of treatment $\E[Y^{a=1} - Y^{a=0}]$ to be equal and, consequently, we do not expect the corresponding identifying observed data functionals, $ \delta_{\phi}$ and $ \delta_{\chi}$, to be equal.

It is useful to observe that under the DAG for Figure \ref{fig:DAG1}A, the identifying functionals for both the per-protocol effect and for the effect of treatment (and their component potential outcome means) involve conditioning on (adjustment for) treatment assignment $Z$. A practical implication of this observation is that under this causal structure, a trial analysis aiming to estimate either of these effects requires information on treatment assignment, in addition to information on the post-assignment covariates. Informally, this is because in this causal structure $Z$ is a confounder of the effect of treatment $A$ on the outcome $Y$, after conditioning on $X$. Specifically, conditioning on $X$ (which is needed because $X$ is a common cause of $A$ and $Y$), opens the $A \leftarrow Z \rightarrow X \leftarrow U \rightarrow Y$ path because $X$ is a collider on that path. 

Readers versed in methods of confounding adjustment will note that the causal structure presented here is one typical of treatment-confounder feedback \cite{robins1986, robins2009, hernan2024causal}. When assignment and treatment are jointly intervened on, conditioning on $X$ is necessary to identify the effect of treatment $A$ on outcome $Y$, but doing so introduces collider stratification bias (sometimes referred to as selection bias) for the effect of assignment $Z$ on $Y$. Furthermore, conditioning on $X$ ``removes'' the part of the effect of assignment $Z$ on the outcome $Y$ that is mediated by $X$. 

%%%%%%%%%%%%%%%%%%%%%%%%%%%%%%%%%%%%%%%%%%%%%%%%%%%%%%%%%%%%%%%%%%%%%%%%%%%%%%
\section*{EXAMPLE CAUSAL STRUCTURE \# 2} \label{section_structure2}
%%%%%%%%%%%%%%%%%%%%%%%%%%%%%%%%%%%%%%%%%%%%%%%%%%%%%%%%%%%%%%%%%%%%%%%%%%%%%%

\begin{figure}[H]
    \caption{Causal structure \#2. (A) Directed acyclic graph; (B) SWIG under intervention to set assignment $Z$ to $z$; (C) SWIG under intervention to set assignment $Z$ to $z$ and treatment $A$ to $a$; and (D) SWIG under intervention to set treatment $A$ to $a$.}
    \centering
    \includegraphics[width=1\linewidth]{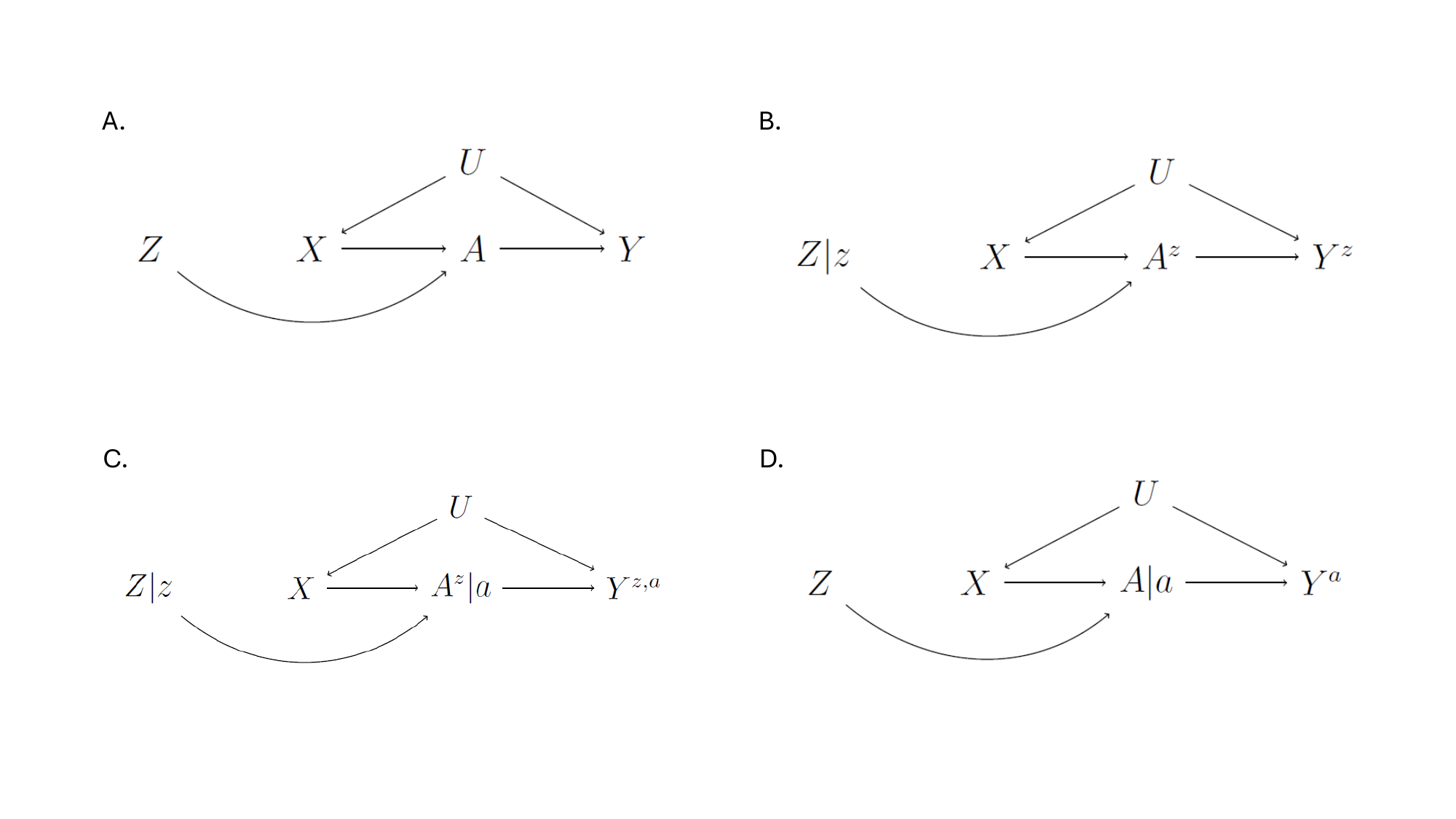}
    \label{fig:DAG2}
\end{figure}

To better appreciate the results in the previous section, it helps to consider a second causal structure obtained from the DAG in Figure \ref{fig:DAG2} by removing the $Z \rightarrow X$ and $X \rightarrow Y$ arrows. The resulting simplified DAG represents an exclusion restriction assumption by eliminating the $Z \rightarrow X \rightarrow Y$ path, such that now assignment $Z$ does not affect the outcome except through treatment $A$. We note in passing that the structure of the DAG in Figure \ref{fig:DAG2}, different from that of the DAG in Figure \ref{fig:DAG1}, does not necessarily impose a temporal ordering between $Z$ and $X$ in the sense that under the DAG in Figure \ref{fig:DAG2}, $X$ could also denote baseline covariates.

The exclusion restriction means that under this causal structure, the counterfactual outcome $Y^{z,a}$ reflecting joint intervention to set assignment $Z$ to $z$ and treatment $A$ to $a$ is equal to the counterfactual outcome $Y^{a}$ reflecting intervention only to set treatment $A$ to $a$, that is, $Y^{z,a} = Y^a$, and we can also conclude that $\E[Y^{z,a}] = \E[Y^{a}]$ \cite{robins1996identification, robins1998correction, hernan2024causal}.

Furthermore, given that the DAG of Figure \ref{fig:DAG2} is obtained by \emph{removing} arrows from the DAG of Figure \ref{fig:DAG1}, it has to be that all independencies that hold under the former remain true under the latter (removal of arrows cannot induce new dependencies). In fact it is easy to check that the independencies used in the previous section are true under the DAG of Figure \ref{fig:DAG2}A by examining the the SWIGs for different interventions. Therefore, we expect all the identification results that we obtained under the DAG of Figure \ref{fig:DAG1} to also hold under the DAG of Figure \ref{fig:DAG2}. It follows that, in the causal structure of Figure \ref{fig:DAG2}A, it remains true that the potential outcome means under joint intervention on assignment and treatment $\E[Y^{z,a}]$ are identified with $\phi(z,a)$ and the potential outcome means under intervention on treatment $\E[Y^{a}]$ are identified with $\chi(a)$; similarly, the per-protocol effect is identified by $\delta_{\phi}$ and the effect of treatment is identified by $\delta_{\chi}$. Furthermore, combining these identification results with the exclusion restriction represented on the DAG of Figure \ref{fig:DAG2}, we can conclude that $\phi(z,a) = \chi(a)$, for $z=0,1$ and $a=0,1$; and $\delta_{\phi} = \delta_{\psi}$. 

Next, we shall argue that under the assumptions embedded in Figure \ref{fig:DAG2}, these identification results can be further simplified. To see this, consider the SWIG obtained from the DAG of Figure \ref{fig:DAG2} for intervention to set treatment $A$ to $a$ (Figure \ref{fig:DAG2}D). In this SWIG, the independence condition $Y^{a} \indep A | X$ holds and therefore, as we show in the Appendix (Sections 1-3), the potential outcome mean under intervention to set treatment $A$ to $a$, $\E[Y^a]$ is identified with $$\psi(a) \equiv \E\big[\E[ Y | X, A = a] \big], \mbox{ with } a = 0,1,$$ and the effect of treatment is identified with $$ \delta_{\psi} = \psi(a=1) - \psi(a=0).$$ 

Combining all these results, we can conclude that under the causal structure of Figure \ref{fig:DAG2}, the potential outcome means under joint intervention on assignment and treatment and under intervention on treatment alone are equal, $\E[Y^{z,a}] = \E[Y^a]$, and identified with $$\phi(z,a) = \chi(a) = \psi(a), \mbox{ for } z = 0,1 \mbox{ and } a = 0,1.$$ Furthermore, the per-protocol effect and the effect of treatment are equal $\E[Y^{z=1,a=1} - Y^{z=0,a=0}] = \E[Y^{a=1} - Y^{a=0}]$, and are identified with  $$\delta_{\phi}=\delta_{\chi}=\delta_{\psi}.$$ 

It is important to note that $\psi(a)$ and $\delta_{\psi}$ do not involve adjustment for (conditioning on) assignment. Therefore, a trial analysis aiming to estimate the effect of treatment (which is equal to the per-protocol effect in this case) under the causal structure of Figure \ref{fig:DAG2} does not require information on assignment to estimate the effect of treatment. Informally, this is because in this causal structure, $Z$ is no longer a confounder of the effect of treatment $A$ on the outcome $Y$, after conditioning on $X$. Furthermore, the equality among different observed data functionals for the effect of treatment, $\delta_{\phi}=\delta_{\chi}=\delta_{\psi}$, and for its component potential outcome means, $\phi(z,a) = \chi(a) = \psi(a)$ for different $z$ and $a$ values, are restrictions on the law (distribution) of the observed data. In other words, our causal assumptions have testable implications that may be useful as falsification tests for some of the identifiability assumptions (see Appendix Section 7 for additional details).

%%%%%%%%%%%%%%%%%%%%%%%%%%%%%%%%%%%%%%%%%%%%%%%%%%%%%%%%%%%%%%%%%%%%%%%%%%%%%%
\section*{REMARKS ON ADDITIONAL CAUSAL STRUCTURES}
%%%%%%%%%%%%%%%%%%%%%%%%%%%%%%%%%%%%%%%%%%%%%%%%%%%%%%%%%%%%%%%%%%%%%%%%%%%%%%

As noted above, the DAG of Figure \ref{fig:DAG2}A was obtained from the DAG of Figure \ref{fig:DAG1}A by removing two arrows. In the Appendix (Sections 1-2), we consider additional DAGs obtained from the DAG of Figure \ref{fig:DAG1}A by removing different arrows. Our examination of these additional causal structures confirms our observations from the causal structures examined in the main text. First, the per-protocol effect and the effect of treatment are not expected to be the same whenever one can trace a directed path from assignment to the outcome that is not mediated by treatment (i.e., when the exclusion restriction does not hold, as was the case in the DAG of Figure \ref{fig:DAG1}A). Second, identification (and estimation) of the per-protocol effect and the effect of treatment require information whenever assignment operates as a confounder of the effect of treatment on the outcome, conditional on other observed covariates included in the analysis. 

In the Appendix (Sections 1-2), we also examine two additional causal structures obtained from the DAG of Figure \ref{fig:DAG1}A; one by removing the $X \rightarrow A$ arrow (see causal structure \#7), and another by removing both the $X \rightarrow A$ and $X \rightarrow Y$ arrows (causal structure \#8). Under these causal structures, we show that expressions other than $\chi(a)$ and $\psi(a)$ can be used to identify the per-protocol effect and the effect of treatment; the two causal estimands are equal between them under the structure of the second graph, but not under the structure of the first graph. Of special interest, in the second of these graphs, the effect of treatment can be identified without adjustment for either assignment $Z$ or the post-assignment variables $X$; however, if investigators decide to condition on $X$, then adjustment for $Z$ becomes necessary for identification. This phenomenon serves as a reminder of the fact that whether a variable (here, assignment $Z$) is a confounder is relative to what other variables are included in the analysis (here, the post-assignment variables $X$) \cite{greenland1986}. 

We note that all results in this paper were obtained for causal structures that do not have a direct effect of assignment on the outcome that was not mediated by observed variables. Adding such a direct effect (e.g., a direct $Z \rightarrow Y$ arrow) to any of the graphs we considered would violate the exclusion restriction and would also render assignment a confounder of the effect of treatment on the outcome, even after conditioning on measured post-baseline variables. Thus, after such modification to any of the causal structures we considered, the per-protocol effect and the effect of treatment would not be equal, and identification (and estimation) of these effects would require conditioning on assignment (in addition to conditioning on the measured post-baseline covariates).

%%%%%%%%%%%%%%%%%%%%%%%%%%%%%%%%%%%%%%%%%%%%%%%%%%%%%%%%%%%%%%%%%%%%
\section*{DISCUSSION}
%%%%%%%%%%%%%%%%%%%%%%%%%%%%%%%%%%%%%%%%%%%%%%%%%%%%%%%%%%%%%%%%%%%%

Using simple causal structures we argued that the per-protocol effect is not necessarily equal to the effect of the treatment strategy. For any randomized trial, these two estimands are equal when assignment only affects the outcome through the treatment -- in these cases one may freely interpret the per-protocol effect as the effect of the treatment strategy. When assignment affects the outcome through paths that do not intersect treatment, then one has to consider interventions both on assignment and on the treatment strategy. 

In other words, our results suggest that the per-protocol effect and the effect of treatment should be considered as distinct estimands whenever assignment has an effect on the outcome not mediated by treatment. When assignment has an effect on outcome not through treatment, investigators need to be clear which effect is of interest to transparently communicate the necessary assumptions and to select appropriate methods for analysis. Furthermore, identification of both the per-protocol effect and the effect of treatment may require information on treatment assignment, unless assignment does not affect the outcome except through treatment and is not a confounder of the treatment-outcome association conditional on other variables in the analysis.

Our results are especially relevant to the interpretation of results from a randomized trial where the mode of treatment assignment (e.g., use of blinding) is materially different from what will be done in clinical practice. Per-protocol effect and treatment effect estimates from such a trial may not apply to clinical practice \cite{dahabreh2019identification} because, when assignment influences the outcome through paths that do not intersect treatment, both effect estimates should be viewed as pertaining to the particular experimental context where the data are generated. The relevance of such results to other populations with different assignment mechanisms should be considered on a case-by-case basis.   

Though we focused on time-fixed treatments, our results extend naturally to trials examining the effects of sustained treatment strategies, including strategies that allow for deferred treatment initiation, treatment discontinuation, or treatment switching when clinically indicated \cite{hernan2017per}. Analyses of such trials require detailed background knowledge about more complex causal structures to determine if assignment only affects the outcome through the treatment strategies of interest. When this strong assumption holds, per-protocol effects are equal to effects of sustained treatment strategies. In these more complex structures, valid identification for both of these estimands also requires careful examination of whether assignment is a confounder of the treatment -- outcome relationship conditional on other variables in the analysis.

%%%%%%%%%%%%%%%%%%%%%%%%%%%%%%%%%%%%%%%%%%%%%%%%%%%%%%%%%%%%%%%%%%%%
\section*{CONCLUSION}
%%%%%%%%%%%%%%%%%%%%%%%%%%%%%%%%%%%%%%%%%%%%%%%%%%%%%%%%%%%%%%%%%%%%

We argued that the definition of per-protocol effects depends on treatment assignment, and that the per-protocol effect is equivalent to the effect of the treatment under the assumption that there are no direct effects of assignment on the outcome. Moreover, we showed that information on assignment may be necessary to identify per-protocol effects, even if assignment is randomized. Our findings suggest the need to more closely consider the distinction between different per-protocol effects (e.g., the effects of different assignment processes) and the effects of treatment strategies in randomized trials.

%%%%%%%%%%%%%%%%%%%%%%%%%%%%%%%%%%%%%%%%%%%%%%%%%%%%%%%%%%%%%%%%%%%%
\section*{ACKNOWLEDGMENTS}
%%%%%%%%%%%%%%%%%%%%%%%%%%%%%%%%%%%%%%%%%%%%%%%%%%%%%%%%%%%%%%%%%%%%

This section has been temporarily removed from the manuscript for peer review.

%%%%%%%%%%%%%%%%%%%%%%%%%%%%%%%%%%%%%%%%%%%%%%%%%%%%%%%%%%%%%%%%%%%%
\clearpage
\section*{CAUSAL STRUCTURES}
%%%%%%%%%%%%%%%%%%%%%%%%%%%%%%%%%%%%%%%%%%%%%%%%%%%%%%%%%%%%%%%%%%%%

\iffalse 

The research reported in this manuscript was supported in part by National Library of Medicine (NLM) award R01LM013616, National Institute of General Medical Sciences (NIGMS) award R35GM154888, and Patient-Centered Outcomes Research Institute (PCORI) awards ME-1502-27794 and ME-2021C2-22365 .\\
The content of this paper is solely the responsibility of the authors and does not necessarily represent the official views of the NLM, NIGMS, PCORI, the PCORI Board of Governors, or the PCORI Methodology Committee.
\fi

%%%%%%%%%%%%%%%%%%%%%%%%%%%%%%%%%%%%%%%%%%%%%%%%%%%%%%%%%%%%%%%%%%%%%%%%%%%%%%
% BIBLIOGRAPHY
%%%%%%%%%%%%%%%%%%%%%%%%%%%%%%%%%%%%%%%%%%%%%%%%%%%%%%%%%%%%%%%%%%%%%%%%%%%%%%
\clearpage
\renewcommand{\refname}{REFERENCES}
\bibliographystyle{ieeetr}
\bibliography{bibliography_cate}

%%%%%%%%%%%%%%%%%%%%%%%%%%%%%%%%%%%%%%%%%%%%%%%%%%%%%%%%%%%%%%%%%%%%%%%%%%%%%%
% VERSIONING 
%%%%%%%%%%%%%%%%%%%%%%%%%%%%%%%%%%%%%%%%%%%%%%%%%%%%%%%%%%%%%%%%%%%%%%%%%%%%%%

\ddmmyyyydate %redefine \today format
\newtimeformat{24h60m60s}{\twodigit{\THEHOUR}.\twodigit{\THEMINUTE}.32}
\settimeformat{24h60m60s}
\begin{center}
\vspace{\fill}\ \newline
\textcolor{black}{{\tiny $ $PP\_effect\_and\_effect\_of\_tx, $ $ }
{\tiny $ $Date: \today~~ \currenttime $ $ }
{\tiny $ $Revision: \paperversionmajor.\paperversionminor $ $ }}
\end{center}

\end{document}